\documentclass[aps,prd,preprint,amsmath,amssymb,nofootinbib]{revtex4}
\usepackage{graphicx}
\def\be{\begin{eqnarray}}
\def\ed{\end{eqnarray}}
\def\ve{\varepsilon}
\def\la{\lambda}
\def\e{\epsilon}

\def\ve{\varepsilon}
\def\la{\langle}
\def\ra{\rangle}
\def\non{\nonumber}
\def\g5{\gamma_{5}}
\def\ga{\gamma}

\begin{document}

\title{ \bf \Large $f_0(1710)$ production in exclusive $B$ decays}

\renewcommand{\thefootnote}{\fnsymbol{footnote}}

\author{\large Chuan-Hung Chen$^{1,2}$\footnote{Email:
{\sf physchen@mail.ncku.edu.tw}} and Tzu-Chiang
Yuan$^{3}$\footnote{Email: {\sf tcyuan@phys.nthu.edu.tw}}
 }

\affiliation{ $^{1}$Department of Physics, National Cheng-Kung
University, Tainan 701, Taiwan \\
$^{2}$National Center for Theoretical Sciences, Taiwan\\
 $^{3}$Department of Physics, National Tsing-Hua University, Hsinchu
300, Taiwan
}

\renewcommand{\thefootnote}{\arabic{footnote}}

\date{\today}

\begin{abstract}
Production of $f_0(1710)$, a theoretical endeavor of pure scalar glueball state, is studied in detail
from exclusive rare $B$ decay within the framework of perturbative QCD.
The branching fraction for $B^\pm \to K^{*\pm} f_0(1710) \to K^{*\pm} (K \bar K)$ is estimated to be
about $8 \times 10^{-6}$, while for $B^\pm \to  K^\pm f_0(1710) \to K^\pm (K \bar K)$ it is smaller by roughly
an order of magnitude.
With the accumulation of almost 1 billion $B \bar B$ pairs from the B{\tiny A}B{\tiny AR}
and Belle experiments to date, 
hunting for a scalar glueball via these rare decay modes should be attainable.

\end{abstract}

\maketitle %

  From the modern point of view, properties of pseudoscalar mesons can
be understood as Nambu-Goldstone bosons due to the spontaneous
symmetry breaking of chiral symmetry. Their low energy dynamics can
be described by the chiral lagrangian. On the other hand, scalar
mesons are not governed by any low energy symmetry like chiral
symmetry and thus they can not take advantage of the power of a
symmetry. Indeed, their $SU(3)$ classification, the quark content of
their composition, as well as their spectroscopy are not well
understood for scalar mesons \cite{scalar-mesons}. Moreover,
possible mixings of the $q \bar q$ states with a pure glueball state
\cite{cheng-chua-liu,he-li-liu-zeng,close-zhao,giacosa-etal,burakovsky-page,
fariborz,chanowitz-talk} must be taken into consideration.

Recent quenched lattice simulation \cite{lattice} predicted the lowest pure glueball state has
a mass equals $1710 \pm 50 \pm 80$ MeV and $J^{PC} = 0^{++}$. The first error is statistical while the second is due to approximate anisotropy of the lattice.
This suggests that before mixing, a glueball mass should be closed to 1710 MeV, instead of the
earlier lattice result of 1500 MeV \cite{cheng-chua-liu}.
This makes  $f_0(1710)$ a strong candidate for a lowest pure glueball state as advocated in
\cite{chanowitz} based on argument of chiral suppression
in $f_0(1710)$ decays into pair of pseudoscalar mesons.
The next two pure glueball states predicted by the quenched approximation  \cite{lattice} have masses
$2390 \pm 30 \pm120$ MeV and $2560 \pm 35 \pm 120$ MeV
with $J^{PC} = 2^{++}$ and $0^{-+}$ respectively.
Mixings between the nearby three isosinglet scalars $f_0(1370)$, $f_0(1500)$, and $f_0(1710)$
and the isovector scalar $a_0(1450)$ have been studied in detail in \cite{cheng-chua-liu} with the following main result:
In the $SU(3)$ symmetry limit, $f_0(1500)$ is a pure $SU(3)$ octet and degenerate with the isovector scalar meson $a_0(1450)$, whereas
$f_0(1370)$ is mainly a $SU(3)$ singlet with a small mixing with $f_0(1710)$ which is composed
predominantly by a scalar glueball.

Important production mechanism of glueballs is the decay of heavy
quarkonium \cite{closeetal, he-jin-ma, zhao-close}.
In fact, the observed enhancement of the mode
$J/\psi \to f_0(1710)\omega$ relative to $f_0(1710)\phi$ and the copious production of
$f_0(1710)$ in the radiative $J/\psi$ decays are strong indication that
$f_0(1710)$ is mainly composed of glueball \cite{cheng-chua-liu}.
Another interesting mechanism is the
direct production from $e^+e^- \to \gamma^* \to  G_J H$ \cite{brodsky},
where $G_J$ stands for a glueball state of spin $J=0$ or 2 and $H$ denotes a $J/\psi$ or
$\Upsilon$.
Recently, glueball production from inclusive rare $B$ decay \cite{he-yuan} has also been studied.
Ironically, scalar glueball state has never been observed in  the gluon-rich channels of $J/\psi(1S)$ decays or $\gamma\gamma$ collisions\footnote{
For a summary of the non-$q\bar q$ candidates from the Particle Data Group, see p949 of
Ref.\cite{pdg}.
}.

\begin{figure}[htbp]
\includegraphics*[width=2.5in]{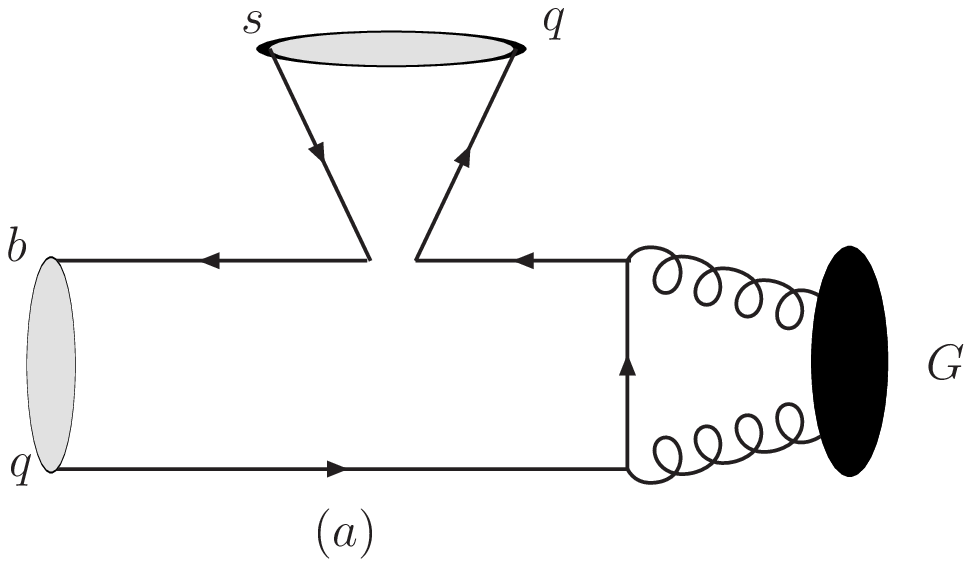}\hspace{0.5cm}\includegraphics*[width=2.5in]{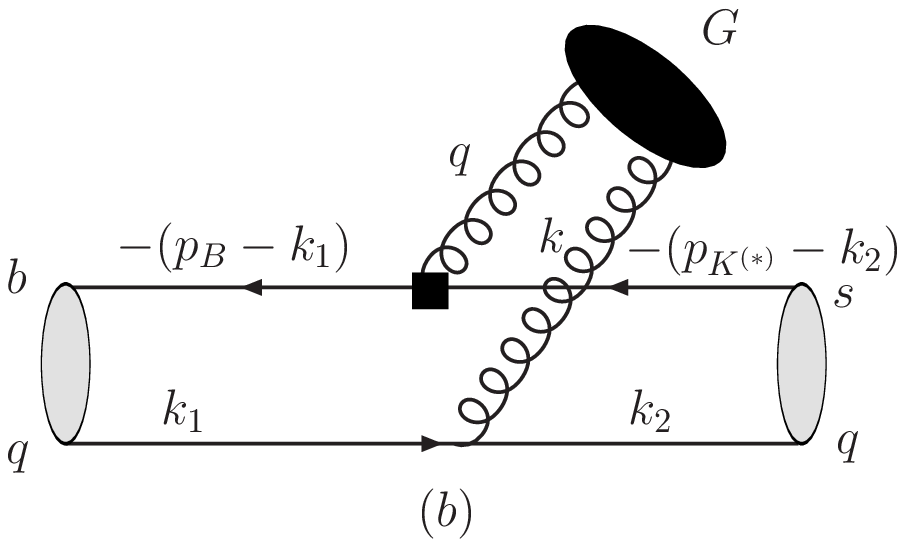}
\caption{Flavor diagrams for the $B\to K^{(*)} {G_0}$.
}
 \label{fig:bkg}
\end{figure}

\begin{figure}[htbp]
\includegraphics*[width=4. in]{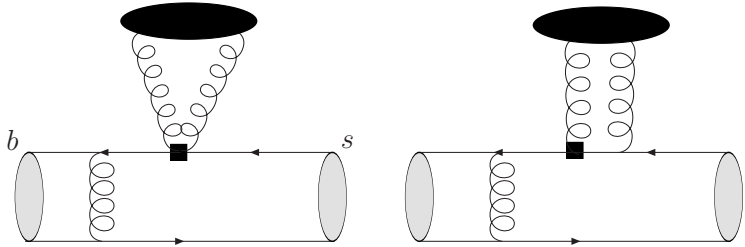}
\caption{Other flavor diagrams for the $B\to K^{(*)} {G_0}$ at higher order in $\alpha_s$.}
 \label{fig:bsgg}
\end{figure}

In this article, we will study glueball production via exclusive $B$
decay using perturbative QCD (PQCD).
Firstly, we will
ignore mixing effects and treat $f_0(1710)$ as a pure scalar
glueball suggested by the quenched lattice data. At the end of
the paper, we will demonstrate the mixing effects are minuscule.
At quark level, the
effective Hamiltonian for the decay $b\to s q \bar{q}$ can be written
as \cite{BBL}
\begin{eqnarray}
 H_{{\rm eff}}&=&{\frac{G_{F}}{\sqrt{2}}}\sum_{q=u,c}V_{q}\left[
C_{1}(\mu) O_{1}^{(q)}(\mu )+C_{2}(\mu )O_{2}^{(q)}(\mu
)+\sum_{i=3}^{10}C_{i}(\mu) O_{i}(\mu )\right] \;,
\label{eq:hamiltonian}
\end{eqnarray}
where $V_{q}=V_{qs}^{*}V_{qb}$ denotes the product of  Cabibbo-Kobayashi-Maskawa
(CKM)  matrix elements  and the operators $O_{1}$-$O_{10}$ are
defined as
\begin{eqnarray}
&&O_{1}^{(q)}=(\bar{s}_{\alpha}q_{\beta})_{V-A}(\bar{q}_{\beta}b_{\alpha})_{V-A}\;,\;\;\;\;\;
\;\;\;O_{2}^{(q)}=(\bar{s}_{\alpha}q_{\alpha})_{V-A}(\bar{q}_{\beta}b_{\beta})_{V-A}\;,
\nonumber \\
&&O_{3}=(\bar{s}_{\alpha}b_{\alpha})_{V-A}\sum_{q}(\bar{q}_{\beta}q_{\beta})_{V-A}\;,\;\;\;
\;O_{4}=(\bar{s}_{\alpha}b_{\beta})_{V-A}\sum_{q}(\bar{q}_{\beta}q_{\alpha})_{V-A}\;,
\nonumber \\
&&O_{5}=(\bar{s}_{\alpha}b_{\alpha})_{V-A}\sum_{q}(\bar{q}_{\beta}q_{\beta})_{V+A}\;,\;\;\;
\;O_{6}=(\bar{s}_{\alpha}b_{\beta})_{V-A}\sum_{q}(\bar{q}_{\beta}q_{\alpha})_{V+A}\;,
\nonumber \\
&&O_{7}=\frac{3}{2}(\bar{s}_{\alpha}b_{\alpha})_{V-A}\sum_{q}e_{q} (\bar{q}%
_{\beta}q_{\beta})_{V+A}\;,\;\;O_{8}=\frac{3}{2}(\bar{s}_{\alpha}b_{\beta})_{V-A}
\sum_{q}e_{q}(\bar{q}_{\beta}q_{\alpha})_{V+A}\;,  \nonumber \\
&&O_{9}=\frac{3}{2}(\bar{s}_{\alpha}b_{\alpha})_{V-A}\sum_{q}e_{q} (\bar{q}%
_{\beta}q_{\beta})_{V-A}\;,\;\;O_{10}=\frac{3}{2}(\bar{s}_{\alpha}b_{\beta})_{V-A}
\sum_{q}e_{q}(\bar{q}_{\beta}q_{\alpha})_{V-A}\;,
\label{eq:operators}
\end{eqnarray}
with $\alpha$ and $\beta$ being the color indices and
$C_{1}$-$C_{10}$ the corresponding Wilson coefficients.
In addition, the gluonic penguin vertex for $b(p) \to s(p') g^*(q)$ with
next-to-leading QCD corrections is given by \cite{he-lin}
\begin{eqnarray}
{\Gamma}^{\mu a} =- \frac{G_F}{\sqrt{2}} \frac{g_s}{4 \pi^2}
V^*_{ts}V_{tb} \bar s(p') \, [ \Delta F_1(q^2 \gamma^\mu - q^\mu
\not\! q) L - i m_b F_2 \sigma^{\mu\nu} q_\nu R ] \, T^a \, b(p),
\label{penguin}
\end{eqnarray}
where $g_{s}$ is the strong coupling constant, $m_b$ is the
$b$-quark mass, $T^a$ is the generator for the color group, and
$L(R)=(1\mp \g5)/2$, $\Delta F_1 = 4\pi(C^{\rm eff}_4(q, \mu) +
C^{\rm eff}_6(q, \mu))/\alpha_s(\mu)$ and $F_2 = -2 C^{\rm
eff}_{8g}(\mu)$.
Explicit formulas for $C^{\rm eff}_{4},C^{\rm eff}_{6},$  and $C^{\rm eff}_{8g}$ can be found in
Ref.~\cite{GFA}. Since the ground state scalar glueball is composed
of two gluons, the effective interaction between a scalar glueball
and gluons can be written as \cite{chanowitz}
 \be
 {\cal L}_{\rm eff}=f_0 G_0 G^{a}_{\mu \nu} G^{a \mu \nu}\,,
 \label{ggglue}
 \ed
where $f_0$ stands for an unknown effective coupling constant, $G_0$
denotes the scalar glueball field, and $G^{a}_{\mu \nu}$ is the gluon field
strength tensor.
With these 4-quarks operators $O_1 - O_{10}$ (\ref{eq:operators}) and the two effective couplings (\ref{penguin}) and
(\ref{ggglue}), we can embark upon the computation of the decay rates for  $B\to
K^{(*)} G_0$ using PQCD.
The flavor diagrams for $B\to K^{(*)} G_0$ decays are
displayed in Fig.~\ref{fig:bkg}.
Fig.(1a) denotes contribution from the 4-quarks operators $O_{1-10}$ given
in Eq.(\ref{eq:operators}),
whereas Fig.(1b) involves the gluonic penguin vertex contribution of Eq.(\ref{penguin}).
Both diagrams are of the same order in $\alpha_s$.
In the heavy quark limit, the production of light meson is supposed to respect color transparency \cite{Bjorken}, {\it i.e.}, final state interactions are subleading
effects and negligible. We will work under this assumption in what follows.
Moreover, diagrams like Fig.\ref{fig:bsgg} that are of higher order in $\alpha_s$ will be ignored.

To deal with the transition matrix elements
for exclusive $B$ decays, we employ PQCD
\cite{PQCD1,PQCD2} factorization formalism to estimate the hadronic
effects. By the factorization theorem, the transition amplitude can
be written as the convolution of hadronic distribution amplitudes
and the hard amplitude of the valence quarks, in which the
distribution amplitudes absorb the infrared divergences and
represent the effects of nonperturbative QCD. As usual, the hard
amplitudes can be calculated perturbatively by following the Feynman
rules. The nonperturbative objects can be described by the nonlocal
matrix elements and are expressed as
\cite{GN_PRD55,K-meson,KLS_PRD65}
 \be
\int \frac{d^4z}{(2\pi)^4}
e^{-i k \cdot z} \la 0 | \bar b_{\beta}
(0) q_{\alpha }(z) | B(p_B) \ra &=& -\frac{i}{\sqrt{2N_c}}
[(\not p_B + m_{B} ) \g5 ]_{\alpha \beta} \phi_{B}(k) \,,
\non\\
\int \frac{d^4z}{(2\pi)^4} e^{-ixp_{K}\cdot z} \la K(p_K) | \bar
q_{\beta }(z) s_{\alpha }(0) | 0 \ra &=& -\frac{i}{\sqrt{2N_c}}
\left\{ [\g5 \not p_K]_{\alpha \beta} \phi_{K}(x) + [\g5]_{\alpha
\beta}m^{0}_{K}
\phi^{p}_{K}(x) \right. \non\\
&& \left.+ m^{0}_{K}[\g5(\not n_{+} \not n_{-} - {\bf 1})]_{\alpha \beta}
\phi^{\sigma}_{K}(x) \right\}\,, \non\\%
\int \frac{d^4z}{(2\pi)^4} e^{-ixp_{K^*}\cdot z} \la
K^*(p_{K^*},\e_L) | \bar q_{\beta }(z) s_{\alpha }(0) | 0 \ra & =&
\frac{1}{\sqrt{2N_c}} \left\{  m_{K^*} [\not\e_L]_{\alpha\beta}
 \phi_{K^*}(x) \right. \non\\
 &&\left.+
[\not\e_L \not p_{K^*}]_{\alpha\beta}
 \phi_{K^*}^p(x)
+ m_{K^*}[{\bf 1}]_{\alpha \beta}
 \phi_{K^*}^{\sigma}(x)
\right\}\,, \label{eq:DA}
 \ed
for $B$, $K$, and  $K^*$ mesons, respectively, where $N_{c}$ is the
number of color, $n_{\pm}$ are two light-like vectors satisfying $n_+ \cdot n_- = 2$,
and $\e_{L}$ is the longitudinal polarization vector
of $K^*$.
$\phi_B(x,b)$, the distribution amplitude of $B$ meson, is constructed
as follows \cite{KLS_PRD65}
\be
\phi_B(x,b) = \int dk^+ d^2k_{\perp} e^{i {\vec k}_{\perp} \cdot {\vec b}} \phi_B(k)
\ed
with $x=k^-/p_B^-$.
$\phi_{K^{(*)}}(x)$ and $\phi^{p,\sigma}_{K^{(*)}}(x)$ are the
twist-2 and 3 distribution amplitudes of $K^{(*)}$ mesons with the argument  $x$ stands for the
momentum fraction.
Finally, $m_B$ and $m_{K^{(*)}}$ are the masses for the $B$ and $K^{(*)}$ with
$m^{0}_{K}=m^2_{K}/(m_s+m_q)$ where $m_q$ and $m_s$ denote the light quark masses.
The meson distribution amplitudes are subjected to the following
normalization conditions
\be
  \int^{1}_{0}
  dx\phi_{B(K^{(*)})}(x)=\frac{f_{B(K^{(*)})}}{2\sqrt{2N_{c}}}\,,\ \ \
\int^{1}_{0}dx
\phi^{p}_{K^{(*)}}(x)=\frac{f^{(T)}_{K^{(*)}}}{2\sqrt{2N_{c}}}\,, \
\ \ \int^{1}_{0}dx \phi^{\sigma}_{K^{(*)}}(x)=0
 \ed
where
$\phi_B (x) = \phi_B(x,0)$ and
$f_{B(K^{(*)})}$ and $f^{(T)}_{K^{(*)}}$ are the decay constants.
We do not introduce transverse momenta for the light mesons $K$ and $K^*$ here which we will
justify later when we discuss the end-point singularities of the decay amplitudes.

In the light-cone coordinate system, we can pick the two light-like vectors to be
 $n_{+}=(1,0,0_{\perp})$ and $n_{-}=(0,1, 0_{\perp})$,
and the momenta of the $B$ and $K$ mesons can be written as
 \be
 p_{B}&=&\frac{m_{B}}{\sqrt{2}}\;(1,1,0_{\perp})\,, \ \ \
p_K=\frac{m_{B}}{\sqrt{2}}(1-r^2_{G_0})(1,0,0_{\perp})\,,
\label{scalar-meson-momenta}
  \ed
 with $r_{{G_0}}=m_{{G_0}}/m_B$.
For the vector meson $K^*$, we take
 \be
p_{K^*}=\frac{m_{B}}{\sqrt{2}}(1-r^2_{G_0},r^2_{K^*},0_{\perp})\,,\
\ \
\e_{L}=\frac{1}{\sqrt{2} \, r_{K^*}}(1-r^2_{{G_0}},-r^2_{K^*},0_{\perp})\,,
\label{vector-meson-momenta}
  \ed
with $r_{K^*}=m_{K^*}/m_B$ in which the physical condition
$\e_{L}\cdot p_{K^*}=0$ is satisfied for massive vector particle.
The momenta of the spectator quarks with their transverse momenta included are given by
 \be
k_{1}=\left(0,\frac{m_B}{\sqrt{2}}x_1,\vec{k}_{1\perp}\right)\,,\ \
\ \ k_{2}=\left(\frac{m_B}{\sqrt{2}}(1-r^2_{G_0})x_2,0,
\vec{k}_{2\perp}\right)\; .
 \ed

With these light-cone coordinates and distribution amplitudes defined,
we can study the transition matrix elements for $B\to M G_0$ ($M=K, K^*$).
We first analyze Fig.~\ref{fig:bkg}(a).
Within the PQCD approach, we find that Fig.~\ref{fig:bkg}(a) is directly proportional to $x_1$.
Since $x_1$ is the momentum fraction
of the valence quark inside the $B$ meson and its value is expected
to be $x_1\approx \bar \Lambda/m_{B}\ll 1$ with $\bar \Lambda =
m_{B}-m_b$.
Comparing to $x_2\sim O(1)$ (Fig.~\ref{fig:bkg}(b)), its contribution
belongs to higher power in heavy quark expansion.
As an illustration, we can use the operator $O_4$ in
Eq.~(\ref{eq:operators}) to demonstrate this effect. Thus, one finds
 \be
 {\cal M}_{O_4}&\propto & \frac{4f_0 g^2_s C_F}{\sqrt{2}}\sqrt{N_c} f_K
 m^6_{B} \int dx_1 dx_2 \frac{d\vec{k}_{1\perp}}{(2\pi)^2} \frac{d\vec{k}_{2\perp}}{(2\pi)^2}
 \left( 1-\frac{m^2_{G_0}}{m^2_{B}}\right)(2-x_2) x_2 x_1 \phi_{B}(x_1,{\vec k}_{1\perp})\non \\
  &\times& \frac{1}{(m^2_{G_0}-m^2_{B} x_1) - |\vec{k}_{1\perp}|^2} \cdot \frac{1}{m^2_{B} x_1 x_2 -
  |\vec{k}_{1\perp}-\vec{k}_{2\perp}|^2} \cdot \frac{1}{m^2_{G_0} (1- x_2) -
  |\vec{k}_{2\perp}|^2} \frac{1}{|k^2_{2\perp}|^2}\,,
  \non\\
  \label{ampl-O-4}
 \ed
where $C_F=(N_c^2-1)/(2N_c)$. It has been shown in \cite{KLS_PRD65}
that under Sudakov suppression arising from $k_\perp$ and threshold
resummations, the average transverse momenta of valence quarks are
$\langle k_{\perp} \rangle \sim 1.5$ GeV and the end point singularities at
$x_{1,2} \to 0$ in Eq.(\ref{ampl-O-4}) can be effectively removed.
With an explicit factor of $x_1$ appearing in the numerator of
Eq.(\ref{ampl-O-4}), this contribution is regarded as a higher power
effect in $1/m_B$ and therefore can be neglected. We note that this
situation is quite similar to the flavor singlet mechanism to the
$B\to \eta^{\prime}$ form factor \cite{BN_NPB651}. According to the
PQCD analysis in Ref.~\cite{Li}, contribution from the possible
gluonic component inside $\eta'$ to the $B\to \eta^{\prime}$ form
factor also has similar behavior. Its numerical value is two orders
of magnitude smaller than the $B\to \pi$ form factor. Similarly,
other operators $O_{1-3}$ and $O_{5-10}$ give the same behavior.
Therefore, to the leading power in $\Lambda_{QCD}/m_{B}$,
the contributions from Fig.~\ref{fig:bkg}(a) can be neglected. We
will concentrate on the contribution of Fig.~\ref{fig:bkg}(b) in
what follows.

By using the introduced nonlocal matrix elements for mesons and the
light-cone coordinates given above, the transition
matrix element for $B\to M G_0$ ($M=K, K^*$) can be obtained from Fig.~\ref{fig:bkg}(b) 
as
 \be
 A_M=\frac{G_F m^{3}_{B}}{\sqrt{2}} V_{ts} V^*_{tb} {\cal M}_{M}
 \label{eq:amp}
 \ed
with the decay amplitude function ${\cal M}_{M}$ given by
 \be
  {\cal M}_{M}&=&\frac{m_{B}}{\pi} f_{0} C_{F} \int^{\infty}_{0} b db\; \int_0^1 dx_1 \int_0^1dx_2
  \phi_{B}(x_1,b) \non\\
  && \times x_{2} \left\{
      e^{(1)}_{M} \phi_{M}(x_2) + e^{(2)}_{M} \phi^{p}_{M}(x_2)
      +e^{(3)}_{M}
      \phi^{\sigma}_{M}(x_2)
  \right\}E(t)h(x_1,x_2,b) \label{eq:ampm}
 \ed
 \be
  e^{(1)}_K&=& \Delta F_1(t) (1- r^2_{G_0}) [1+2r^2_{G_0} +2(1-r^2_{G_0}) x_2]-3r_b(1-r^2_{G_0})
  F_2(t)
  \,, \non\\
  e^{(2)}_{K} &=& 3 r_K [-2\Delta F_1(t) (r^2_{G_0}+(1-r^2_{G_0})x_2)+r_b F_2(t)
  (1+r^2_{G_0}+(1-r^2_{G_0})x_2)]\,, \non\\
   e^{(3)}_{K}&=& r_b r_K (1-r^2_{G_0})(1-x_2)F_2(t)\,,
   \label{e123}
\ed for the pseudoscalar $K$, and \be
  e^{(1)}_{K^*}&=& e^{(1)}_{K}\,, \ \ \
  e^{(2)}_{K^*}= \frac{r_{K^*}}{r_{K}} (1 - r^2_{G_0} ) e^{(3)}_{K} \,,\ \ \
  e^{(3)}_{K^*}= \frac{r_{K^*}}{r_{K}} e^{(2)}_{K} \,,
 \ed
for the vector meson $K^*$. Here we have introduced the
dimensionless variables $r_b=m_b/m_B$, $r_{K}=m^0_{K}/m_B$, and
$r_{K^*}=m_{K^*}/m_B$.
The hard function $h(x_1,x_2,b)$ in Eq.(\ref{eq:ampm}) is given by
   \be
    h(x_1,x_2,b)&=& \frac{1}{X_{12}+Y_{12}}\left[K_{0}\left(\sqrt{m^2_B Y_{12}}\;b\right)
    - i \frac{\pi}{2} H^{(1)}_{0}\left(\sqrt{m^2_B X_{12}}\;b\right) \right]
   \ed
with $X_{12}=(1-x_1) [r^2_{G_0}+(1-r^2_{G_0})x_2]$ and
$Y_{12}=(1-r^2_{G_0})x_1 x_2$.
The evolution factor $E(t)$ in Eq.(\ref{eq:ampm})
is defined by
  \be
   E(t)= \alpha_{s}(t)e^{-S_B(t)-S_{K}(t)}\,,
   \label{evolution}
  \ed
where $\exp(-S_{B(K)})$ is the Sudakov exponents  that resummed
large logarithmic corrections to the $B (K)$ meson wave
functions \cite{CL_PRD63,LL_PRD70}. Their explicit forms are given by
 \be
 S_B(t)=s(x_1 p^{+}_B,b)+\frac{5}{3}\int^{t}_{1/b} \frac{d\bar \mu}{\bar
 \mu}\ga(\alpha_s(\bar \mu))\,, \non\\
  S_K(t)=s(x_2 p^{+},b)+s((1-x_2) p^{+},b)+2 \int^{t}_{1/b} \frac{d\bar \mu}{\bar
 \mu}\ga(\alpha_s(\bar \mu))\,,
 \label{sudakov}
 \ed
where $\ga(\alpha_s(\mu))$ is the anomalous dimension.
To leading order in $\alpha_s$,  $\ga(\alpha_s(\mu))$ equals $-\alpha_s/\pi$.
The function $s(Q,b)$ in Eq.(\ref{sudakov}) is given by \cite{CS,BS}
 \be
 s(Q,b)=\int^{Q}_{1/b} \frac{d\mu}{\mu}\left[ \ln\left( \frac{Q}{\mu}\right)A(\alpha_s(\mu))+
 B(\alpha_s(\mu))\right]\,,
 \ed
 where
 \be
A &=& C_F \frac{\alpha_s}{\pi} + \left[
\frac{67}{9}-\frac{\pi^2}{3}-\frac{10}{27}f+\frac{2}{3} \beta_0
\ln\left(
\frac{e^{\ga_E}}{2}\right)\right]\left(\frac{\alpha_s}{\pi}\right)^2\,,
\non \\
B&=&\frac{2}{3}\left(
\frac{\alpha_s}{\pi}\right)\ln\left(\frac{e^{2\ga_E-1}}{2} \right)
  \ed
with $f=4$ being the active flavor number and $\ga_E$ is  the Euler
constant.
As mentioned before, $x_1\approx \bar \Lambda/m_{B}\ll 1$,
we have dropped all terms related to $x_1$ in the
above expressions for $\{e^{(i)}_{M} \}$. Since $r_{K^{(*)}} \ll 1$,
we have retained only those terms in the above formulas for
$\{e^{(i)}_{M} \}$ that are at most linear in $r_{K^{(*)}}$. The
scale $t$ where the strong coupling $\alpha_s(t)$ in (\ref{evolution}),
the Sudakov exponents in (\ref{sudakov}),
and the $\Delta F_1(t)$ and $F_2(t)$  in (\ref{e123}) are evaluated
will be discussed later.
For comparison, we also present the formula
of the decay amplitude function ${\cal M}_M$ with $k_{\perp} = 0$ in Appendix~\ref{ap2}.

For estimating our numerical results, we take the values of
theoretical parameters to be: $f_{B}=190$ MeV, $m_b=4.4$ GeV,
$(m_{B},\, m_K,\, m_{K^*},\, m_{G_0})=(5.28,\, 0.493,\, 0.892,
1.71)$ GeV, $V_{ts}V^*_{tb}=-0.041$. For the $B$ meson distribution amplitude, we
use \cite{CL_PRD63}
 \begin{eqnarray}
   \phi_{B}(x_1,b)=N_{B}x_1^2(1-x_1)^2\exp\left[
   -\frac{1}{2} \left( \frac{m_{B} x_1}{\omega_B} \right)^2
   -\frac{1}{2} \omega_B^2 b^2
   \right]
 \end{eqnarray}
with $N_{B}=111.2$ GeV and $\omega_B=0.38$ GeV.
For the distribution amplitudes of the light
pseudoscalar $K$ and vector mesons $K^*$, we refer to their results derived by
the light-cone QCD sum rules in \cite{BBL_JHEP,BZ_PRD71,BBKT}. Their
explicit expressions and relevant values of parameters are collected
in the Appendix~\ref{DA} for convenience.

According to the results
of light-cone QCD sum rules, at small $x_2$, the behavior of twist-2
distribution amplitude obeys the asymptotic form
$\phi_{K^{(*)}}(x_2) \propto  x_2(1-x_2)$, whilst those of twist-3
distribution amplitudes approach a constant
$\phi^{p,\sigma}_{K^{(*)}} (x_2) \propto {\rm const}$. Consequently,
at small $x_2$, the decay amplitude function contributed by the
twist-2 distribution amplitudes of $K^{(*)}$  behaves like
 \be
{\cal M}^{\rm tw2}_{K^{(*)}}\propto  \frac{x_{2} \phi_{B}(x_1) \phi_{K^{(*)}}(x_2)}{k^2
q^2} \propto  \frac{ x_{2} x^2_1 (1-x_1)^2 x_2(1-x_2)}{x_1 x_2
(r^2_{G_0} + (1-r^2_{G_0})x_2)} = \frac{x_1 (1-x_1)^2 x_2
(1-x_2)}{(r^2_{G_0} + (1-r^2_{G_0})x_2)}\,.
 \ed
Obviously, even if one sets $r_{G_0}$ to be zero, the effects from
twist-2 distribution amplitudes of $K^{(*)}$ are well-defined  at the end point $x_2 \to 0$.
Similarly, the contribution from twist-3 distribution amplitudes to the decay amplitude function
at small $x_2$ behaves like
 \be
{\cal M}^{\rm tw3}_{K^{(*)}}\propto \frac{x_{2} \phi_{B}(x_1)
\phi^{p,\sigma}_{K^{(*)}}(x_2)}{k^2 q^2} \propto \frac{ x_{2} x^2_1
(1-x_1)^2 }{x_1 x_2  (r^2_{G_0} + (1-r^2_{G_0})x_2)} = \frac{x_1
(1-x_1)^2 }{(r^2_{G_0} + (1-r^2_{G_0})x_2)}\,.
 \ed
Whence $r_{G_0} \to 0$, one will suffer logarithmic
divergences from the twist-3 distribution amplitudes. In practice,
$r_{G_0} \sim 0.32$, the divergence will not occur. This implies that the influence of
$k_{\perp}$ can only be mild. As a common practice, we do not introduce
transverse momenta for the valence quarks to suppress large effects from end
point singularities.

Since the Wilson coefficients are $\mu$ scale dependence, for
smearing its dependence, we include the values of Wilson
coefficients with the next-to-leading QCD corrections \cite{GFA}.
However, even so, the $C^{\rm eff}_{4,6,8g}$ are still slightly
$\mu$-dependence. Due to this reason, determination of the scale of
exchanged hard gluons in Fig.~\ref{fig:bkg} is also one of the
origins of theoretical uncertainties. For the gluon that attached to
the penguin vertex $b\to s g^*$,  it carries a typical momentum of
$\sqrt{q^2}=m_B \left( (1-x_1)(r^2_{G_0}+(1-r^2_{G_0}) x_2
\right)^{1/2}$. When $x_1 \sim {\bar \Lambda} /m_b$ and $x_2$ is
$O(1)$, say $x_2=0.5$, we get $\sqrt{q^2}\sim 3.9$ GeV. However, the
gluon attached to the spectator quark carries roughly a typical
momentum of $\sqrt{-k^2}=m_B\left( (1-r^2_{G_0})x_1 x_2
\right)^{1/2} \sim 1.4 $ GeV. We note that a suitable range of $x_2$
in PQCD is often taken as $\sim 0.3-0.7$. For definiteness, we take
the democratic average value $t=(\sqrt{q^2}+\sqrt{-k^2})/2$ as the
hard scale, in which the allowed value is within the range $t\approx
2.45 \pm 0.45$ GeV. This justifies somewhat the validity of the PQCD
approach and we will take this range of $t$ as our theoretical
uncertainties. For illustration, we present the involving Wilson
coefficients at different values of $\mu$ scale in
Table~\ref{tab:WC}.
\begin{table}[hptb]
\caption{ The involving Wilson coefficients at various values of $\mu$ scale.
}\label{tab:WC}
\begin{ruledtabular}
\begin{tabular}{cccc}
Wilson coefficient & $\mu=2.1$ GeV& 2.5 GeV & 3.0 GeV
 \\ \hline
$C^{\rm eff}_4$ & $-(6.17+0.78i) \times 10^{-2}$ & $-(5.80+0.89i)\times 10^{-2}$ & $-(5.48+ 0.89i)\times 10^{-2}$ \\
\hline
$C^{\rm eff}_6$ & $-(7.69+ 0.78i) \times 10^{-2}$ & $-(7.19+0.89i)\times
10^{-2}$ & $-(6.77+0.89i)\times 10^{-2}$\\ \hline
 $C^{\rm eff}_{8g}$ &  $-0.170$ & $-0.165$  & $-0.161$  \\
\end{tabular}
\end{ruledtabular}
\end{table}

\begin{figure}[htbp]
\includegraphics*[width=2.5in]{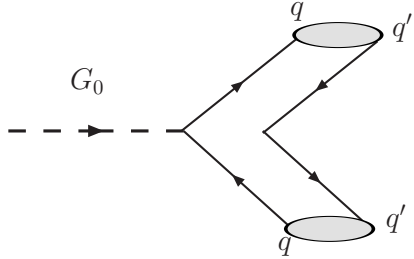}
\caption{Flavor diagram for the $G_0\to P P$ with $P$ being a
pseudoscalar. }
 \label{fig:G0qq}
\end{figure}

Effective interactions between a scalar glueball $G_0$ and the pseudoscalars
have been studied using chiral perturbation theory \cite{chao-he-ma,he-yuan}.
By using the current experimental data \cite{pdg}
$\Gamma_{\rm total} (f_0(1710))=137 \pm 8$ MeV and BR$(f_0(1710)\to K\bar
K)=0.38^{+0.09}_{-0.19}$,
this allows us to get an estimate of the unknown coupling $f_0=0.07^{+0.009}_{-0.018}$
GeV$^{-1}$ \cite{he-yuan}.
This result of $f_0$ should be taken as a crude estimation. For one thing,
the data of the branching ratio BR$(f_0(1710)\to K\bar
K)$ was not used for averages, fits, limits, etc. by the PDG \cite{pdg}.
Instead the following two ratios were used in the PDG analysis:
\begin{eqnarray}
R_{\eta/K} \equiv \frac{\Gamma(f_0(1710) \to \eta\eta)}{\Gamma(f_0(1710) \to K {\bar K})} & = & 0.48 \pm 0.15 \; ,
\label{etaetatokk}\\
R_{\pi / K} \equiv \frac{\Gamma(f_0(1710) \to \pi\pi)}{\Gamma(f_0(1710) \to K {\bar K})} & < & 0.11 \; .
\label{pipitokk}
\end{eqnarray}
Within the approach of chiral perturbation theory \cite{he-yuan}, it would be difficult to accommodate these two ratios of Eqs.(\ref{etaetatokk}) and (\ref{pipitokk}), since the leading term in the chiral Lagrangian is flavor blind. Here we will present another approach to estimate $f_0$.
At quark level, the amplitude for $G_{0}\to q\bar q$ is proportional to the
quark mass $m_q$ and therefore chirally suppressed. Its explicit form is 
given by \cite{chanowitz}
 \be
  A(G_0\to q\bar q)&=& -f_0 \alpha_s \frac{16\pi \sqrt{2}}{3}
  \frac{m_q}{\beta} \ln \left( \frac{1+\beta}{1-\beta} \right) \bar u_{q} v_{q}\,,
  \label{eq:g0qq}
 \ed
where $\beta$ denotes the velocity of the quark and $u_q\,(v_q)$ is the
quark (anti-quark) spinor. 
It has been argued in \cite{chanowitz} that the chiral suppression 
of the amplitude $A(G_0\to q\bar q) \propto m_q$ persist in all order of $\alpha_s$.
One may treat the coefficient of this decay amplitude
as the short-distance coefficient of the strong decay  $G_0 \to PP$ where $P$ stands for
a pseudoscalar meson like $\pi$, $K$, or $\eta $ etc, as illustrated in Fig.~\ref{fig:G0qq}.
Thus,
 \be
  \la PP | {\cal H}_{\rm eff}| G_0\ra &=& -f_0 m_q Y F^{PP}(m^2_{G_0})  \,,
  \label{eq:ampGPP}
 \ed
 with, to leading order in $\alpha_s$,
 \be
  Y&=& \alpha_s ( m^2_{G_0} )\frac{16\pi \sqrt{2}}{3}
  \frac{1}{\beta} \ln \left( \frac{1+\beta}{1-\beta} \right) \, ,
 \ed
 and $F^{PP}(m^2_{G_0})$ is the time-like form factor $\la PP |\bar q q|0 \ra$
 evaluated at $Q^2=m^2_{G_0}$.
For the case of $P=\eta$, we include the quark-flavor mixing effect
according to  $\eta= \cos\phi
\,\eta_q-\sin\phi \, \eta_s$ and $\eta'=\sin \phi \, \eta_q+\cos\phi \, \eta_s$ with
$\eta _{q}  = ( {u\bar u + d\bar d})/\sqrt{2}$, $\eta_{s} = s\bar s
$ \cite{flavor0,flavor}, and $\phi=41.4^{\circ}$ \cite{KLOE}.
Using Eq.~(\ref{eq:ampGPP}), the following ratios of the partial decay rates can be obtained
 \be
 R_{\pi/K}&=&\frac{\Gamma(G_0\to \pi \pi)}{\Gamma(G_0\to K\bar K)}=
\frac{3}{8} \frac{(m_u+m_d)^2}{m^2_s}
\frac{|F^{\pi\pi}(m^2_{G_0})|^2}{|F^{KK}(m^2_{G_0})|^2}
\frac{\left(1-4m^2_{\pi}/m^2_{G_0} \right)^{1/2}}
{\left(1-4m^2_{K}/m^2_{G_0}
\right)^{1/2}}\,, \non\\
R_{\eta/K}&=&\frac{\Gamma(G_0\to \eta \eta)}{ \Gamma(G_0\to K\bar
K)}= \frac{\left(1-4m^2_{\eta}/m^2_{G_0} \right)^{1/2}}
{\left(1-4m^2_{K}/m^2_{G_0} \right)^{1/2}} \non
\\
&\times& \frac{ |(m_u+m_d)\cos^2\phi F^{\eta_q \eta_q}(m^2_{G_0})/2+
m_s \sin^2\phi F^{\eta_s\eta_s}(m^2_{G_0})|^2}
{ 2 m^2_s |F^{KK}(m^2_{G_0})|^2} \,.
 \ed
By taking the flavor $SU(3)$ approximation, one finds that $F^{\pi
\pi}/F^{KK}\approx f^2_{\pi}/f^2_K$, $F^{\eta_q
\eta_q}/F^{KK}\approx f^2_{q}/f^2_{K}$, and
$F^{\eta_s\eta_s}/F^{KK}\approx f^2_s/f^2_K$. With $m_u=m_d=10$ ,
$m_s=120$, $f_{\pi}=130$, $f_{K}=160$ \cite{pdg}, $f_q=140$,
$f_s=180$ \cite{flavor} (all in unit of MeV), one deduces
 \be
R_{\pi/K}= 0.006\,, \
 \ \ \
R_{\eta/K}= 0.37\,.
 \ed
Identifying $G_0$ to be $f_0(1710)$, these ratios
are consistent with the current experimental data
quoted in Eqs.(\ref{etaetatokk}) and (\ref{pipitokk}).
Using Eq.~(\ref{eq:g0qq}),  the following expression of $f_0$ can be
obtained
 \be
  f^2_{0}&=& \frac{8\pi m_{G_0} \Gamma_{G_0}}{|F^{KK}(m^2_{G_0}) m_s Y
  |^2} \left( 1-\frac{4m^2_{K}}{m^2_{G_0}}\right)^{1/2} {\rm BR}(G_0\to KK) \; ,
 \ed
where $\Gamma_{G_0}=137\pm 8$ MeV is identified as the width of $f_0(1710)$.
The time-like form factor $F^{KK}(m^2_{G_0})$ has been extracted in Ref.~\cite{CCS}
by performing the data fitting to non-resonant $B\to KKK$ decays with the following form
 \be
  F^{KK}(Q)&=& \frac{v}{3} \left( 3F^{(1)}_{NR}+ 2F^{(2)}_{NR}\right)
  + \sigma_{NR} \exp(-\alpha_{NR} Q^2)\,,\non\\
  F^{(n)}_{NR}&=&
\left(\frac{x^{(n)}_{1}}{Q^2}+\frac{x^{(n)}_{2}}{Q^4} \right)
\left( \ln \frac{Q^2}{\Lambda^2}\right)^{-1}\,,
 \ed
where $v=(m^2_{K}-m^2_{\pi})/(m_s-m_d)$, $\Lambda=0.3$ GeV,
$x^{(1)}_{1}=-3.26$ GeV$^2$, $x^{(1)}_{2}=5.02$ GeV$^4$,
$x^{(2)}_{1}=0.47$ GeV$^2$, $x^{(2)}_{2}=0$, $\sigma_{NR}=4.4
e^{i\pi/4}$ GeV, and $\alpha_{NR}=0.13$ GeV$^{-2}$. By using
BR$(G_0\to KK)=0.38^{+0.09}_{-0.19}$ \cite{pdg}, the value for $f_0$ is estimated to be $f_0=0.086^{+0.010}_{-0.026}$,
which is only slightly larger than the value obtained from the chiral 
Lagrangian approach. 
In passing, we note that, using light-cone distribution amplitudes, it has been argued in Ref.\cite{chao-he-ma} that $G_0 \to \pi\pi, K \bar K$ might be dominated by the 4-quark process of $G_0 \to q \bar q  q \bar q$ which is not chirally suppressed. Using this 4-quark mechanism and PQCD factorization scheme, one would predict a large ratio of $R_{\pi/K} \approx (f_\pi /f_K)^4 \approx 0.48$. For further discussion of this issue, we refer our reader to Refs.\cite{chao-he-ma,chao-he-ma-comment,chanowitz-reply}.

\begin{table}[hptb]
\caption{Decay amplitude ${\cal M}_{M}$ (in units of $10^{-4}$) for
$B^{+}\to (K^{+},\, K^{*+}) G_0 $  with and without $k_{\perp}$ at
$f_{0}=0.086$ GeV$^{-1}$ and three different choices of $\mu=2.1$,
$2.5$, and $3.0$ GeV. Numbers given in brackets are without
$k_\perp$. } \label{tab:ampm}
\begin{ruledtabular}
\begin{tabular}{cccc}
Mode  & $\mu=2.1$ GeV & 2.5 GeV & 3.0 GeV
 \\ \hline
 $K^+ G_0$  & $-3.54- 0.42 i$ & $-3.34 -  0.44i$ & $-3.22- 0.48i$ \\
   & $(-3.51- 0.38 i)$ & $(-3.28 -  0.41i)$ & $(-3.08- 0.43i)$ \\ \hline
 $K^{*+} G_0$ & $-11.13-1.51i$  &  $-12.56 -  1.51i$  &  $-12.40 -
 1.70 i$ \\
  & $(-10.90-1.17i)$  &  $(-10.18 -  1.25i)$  &  $(-9.60 - 1.33i)$ \\
\end{tabular}
\end{ruledtabular}
\end{table}
Using the matrix element defined by Eq.~(\ref{eq:ampm}) with the
above chosen values of parameters, the values of ${\cal
M}_{K^{(*)}}$ are given in Table~\ref{tab:ampm} for $f_{0}=0.086$
GeV$^{-1}$ and three different values of $\mu$ scale. For
comparisons, we also present the results with
$k_{\perp} = 0$ in  Table~\ref{tab:ampm}.

The branching fractions for $B^{+}\to
(K^{+},\, K^{*+}) G_0$ decays are tabulated in Table~\ref{tab:Br}.
  From Table~\ref{tab:Br}, we find that the branching fraction for
$B^{+}\to K^{*+} G_0$ is about one order of magnitude larger than
that for $B^{+}\to K^+ G_0$. The difference arises not only from the
values of the decay constants $f_{K}$ and $f_{K^*}$ entered in the
distribution amplitudes, but also from the effects of
$e^{(2)}_{K}\phi^{p}_{K}(x_2)$ and
$e^{(3)}_{K}\phi^{\sigma}_{K}(x_2)$ in the $K^+ G_0$ mode, which are
switched to $e^{(2)}_{K}\phi^{\sigma}_{K^*}(x_2)$ and
$e^{(3)}_{K}\phi^{p}_{K^*}(x_2)$ respectively in the $K^{*+} G_0$
mode.
We also find that the $k_{\perp}$
influence on $B^{+}\to K^{*+} G_0$ decay is stronger than $B^{+}\to
K^{+} G_0$. In addition, when $\mu$ is smaller, $k_{\perp}$ has lesser effects on
the decay $B^{+}\to K^{*+} G_0$. 
\begin{table}[hptb]
\caption{ Branching fractions (in units of $10^{-6}$) for $B^{+}\to
(K^{+},\, K^{*+}) G_0 $  with  and without $k_{\perp}$ at
$f_{0}=0.086$ GeV$^{-1}$ and three different choices of  $\mu=2.1$,
$2.5$, and $3.0$ GeV. Numbers given in brackets are without
$k_\perp$.} \label{tab:Br}
\begin{ruledtabular}
\begin{tabular}{cccc}
Mode  & $\mu=2.1$ GeV & 2.5 GeV & 3.0 GeV
 \\ \hline
 $K^+ G_0$  & $3.05$ & $2.72$ & $2.53$ \\
  & $(2.99)$ & $(2.62)$ & $(2.34)$ \\ \hline
 $K^{*+} G_0$ & $29.07$  &  $35.94$  &  $36.06$ \\
  & $(26.50)$  &  $(23.21)$  &  $(20.69)$ \\
\end{tabular}
\end{ruledtabular}
\end{table}

The branching fractions for the decay chains
$B^{+}\to K^{+} G_0\to K^{+} (K\bar K)_{G_0}$ and $B^{+}\to K^{*+}
G_0\to K^{+} (K\bar K)_{G_0}$ are tabulated in Table~\ref{tab:Br_f},
where the errors are coming from the experimental data of
BR$(f_{0}(1710)\to K\bar K)$.  From Table~\ref{tab:Br_f}, we learn
that one has a better chance to look for the ground state of
glueball through the three-body decays $B\to K^* K \bar K$, since
its branching fraction could be more than a factor of 10 larger than
$B\to K K \bar K$.
\begin{table}[hptb]
\caption{ Branching fractions (in units of $10^{-6}$) for $B^{+}\to (K^{+},\,
K^{*+}) (K\bar K)_{G_0} $
at $\mu=2.1$, $2.5$, and $3.0$ GeV.  Numbers given in brackets are without
$k_\perp$. }
\label{tab:Br_f}
\begin{ruledtabular}
\begin{tabular}{cccc}
Mode  & $\mu=2.1$ GeV & 2.5 GeV & 3.0 GeV
 \\ \hline
 $K^+ (K\bar K)_{G_0}$  & $1.16^{+0.63}_{-0.88}$ &
 $1.03^{+0.56}_{-0.78}$
 & $0.96^{+0.52}_{-0.73}$ \\
 & $\left(1.13^{+0.62}_{-0.85}\right)$ &
 $\left(1.00^{+0.53}_{-0.76}\right)$
 & $\left(0.89^{+0.48}_{-0.67}\right)$ \\

 \hline
 $K^{*+} (K \bar K)_{G_0}$ & $11.05^{+5.98}_{-8.36}$  &  $13.66^{+7.39}_{-10.34}$
 &  $13.70^{+7.42}_{-10.37}$ \\
  & $\left( 10.07^{+5.45}_{-7.62}\right)$  &  $\left(8.81^{+4.77}_{-6.66}\right)$
 &  $\left(7.86^{+4.26}_{-5.95}\right)$
\end{tabular}
\end{ruledtabular}
\end{table}
%
Recently, B{\tiny A}B{\tiny AR} had reported the following branching ratio for $B^\pm \to (K^+K^-)K^\pm$ where 
the $(K^+K^-)$ pair coming from the $f_0(1710)$ 
\cite{babar-br} 
\be
{\rm BR}(B^\pm \to (K^+K^-)_{f_0(1710)} K^\pm) = (1.7 \pm 1.0 \pm 0.3)  \times 10^{-6}\; .
\ed
 From the first and second rows in Table \ref{tab:Br_f}, identifying $G_0$ to be $f_0(1710)$, one can see that our predictions are consistent with the experimental data.

\begin{figure}[htbp]
\includegraphics*[width=4 in]{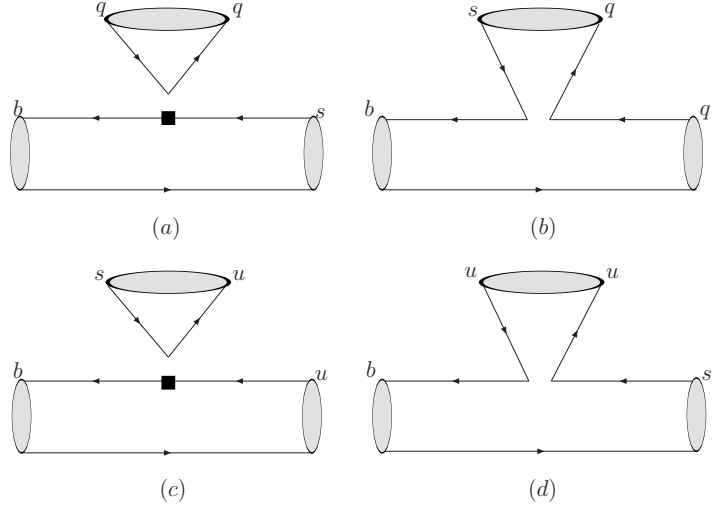}
\caption{Flavor diagrams for the $B\to K^{(*)} (N, S)$. (a) and (b)
are from QCD and electroweak penguin diagrams, while (c) and (d)
denote the tree contributions.
}
 \label{fig:bkqq}
\end{figure}

Before we close, we want to address the issue of mixing effects.
Although we have treated $f_0(1710)$ as a pure
gluonic state, it should be interesting to consider its mixing
effects with other $q\bar q$ states.
To deal with the mixture of a pure glueball with the $q\bar q$ quarkonia state,
we follow Ref.~\cite{cheng-chua-liu} to express the $f_0(1710)$ state 
as the following combination
 \be
 |f_0(1710)\rangle= C_N |N \rangle + C_S | S \rangle + C_G | G\rangle
 \ed
 where $\vert G \rangle$ is the pure glueball state,
 $|N\rangle= (u\bar u+ d\bar d)/\sqrt{2}$, and $|S\rangle=s\bar s$.
Accordingly to one of the mixing schemes \cite{cheng-chua-liu},
the coefficients took the following values:
$C_N=0.32$, $C_S=0.18$, and $C_G=0.93$.
The quoted results of these coefficients are similar
to those obtained by others in  Refs.~\cite{burakovsky-page,LW}. The
corresponding flavor diagrams for the decays $B\to K^{(*)} (N,\, S)$
are shown in Fig.~\ref{fig:bkqq}.
Since the distribution amplitude and decay constant for $f_0(1710)$
are uncertain,  for simplicity, we use factorization
assumption to estimate the hadronic effects for these two-body $B$
decays. In terms of the operators in Eq.~(\ref{eq:operators}),
one can easily show that the contributions from diagram
Fig.~\ref{fig:bkqq}(a) and (d) are associated with the matrix element
$\langle N(S)|\bar q \gamma_{\mu} q|0\rangle$.
Since the scalar $N$ or $S$ is $C$-even while $\bar q \gamma_{\mu} q$ is $C$-odd,
the contributions from Fig.~\ref{fig:bkqq}(a) and (d) must vanish
because charge conjugation is a good quantum number in strong interaction.
On the other hand, the contributions from Fig.~\ref{fig:bkqq}(b) and (c) are non-vanishing and they can be derived as
 \be
  A_{KN}&=&\frac{G_F}{\sqrt{2}} \frac{f_{K} C_{N}}{\sqrt{2}} (m^2_{B} -
  m^2_{N})\left[ V_{ts} V^*_{tb} \left( a^{u}_{4}-\rho_{K} a^u_6
  \right) -V_{us} V^*_{ub} \, a_{1} \right] F^{BN}_{0}(m^2_{K})\,, \\
A_{KS}&=&-\frac{G_F}{\sqrt{2}} V_{ts}V^{*}_{tb} \left[ 2 m_S f_S C_S
\frac{m^2_B-m^2_K}{m_b+m_s} a^{s}_6\right]
  F^{BK}_0(m^2_S)\,, \\
 A_{K^* N}&=&\frac{G_F}{\sqrt{2}}  \frac{f_{K^*} C_{N}}{\sqrt{2}} (m^2_{B} -
  m^2_{N}) \left[ V_{ts} V^*_{tb} \, a^{u}_{4}
    - V_{us} V^*_{ub} \,a_1\right]F^{BN}_{0}(m^2_{K^*})\,, \\
A_{K^*S}&=& \frac{G_F}{\sqrt{2}} V_{ts}V^{*}_{tb} \left[ 2 m_S f_S
C_S \frac{m^2_B}{m_b-m_s} a^{s}_{6}\right] A^{BK^*}_{0} (m^2_{S})
     \label{eq:ampqq}
 \ed
for $B^{+}\to K^{+} (N,\, S)$ and $B^{+}\to K^{*+} (N,\, S)$ decays,
respectively,
where $\rho_K$, $a_1$, and $a^{u}_{4,6}$ are defined by
 \be
    \rho_{K}&=& \frac{2\, m^2_{K}}{(m_s+m_u)(m_b+m_u)}\,, \non\\
    a_1&=& C_2 + \frac{C_1}{\sqrt{N_c}}\,, \non\\
    a^{q}_{4}&=&C_{4}+\frac{C_3}{N_{c}}+\frac{3}{2}e_{q}\left(C_{10}
        +\frac{C_{9}}{N_{c}}\right)\,,
  \non \\
     a^{q}_{6}&=&C_{6}+\frac{C_5}{N_{c}}+\frac{3}{2}e_{q}\left(C_{8}+\frac{C_{7}}{N_{c}}\right)\,.
  \ed
$e_q$ is the electric charge of quark $q$ and $F^{BM}_0$ with
$M=N,\, S$ and $A^{BK^*}_0$ correspond to the $B\to (M,\, K^*)$ form
factors parametrized by \cite{CG_PRD75,LF1}
  \be
 \la  N(p)| \bar b \ga_{\mu}\ga_5 q|B(p_{B})\ra &=&
 -i\left[\left(P_\mu- \frac{m^{2}_{B}-m^{2}_{N}}{q^2}\,q_
 \mu\right) F_1^{BN}(q^2) \right.
 \left.+ \frac{m^{2}_{B}-m^{2}_{N}}{q^2}
 \,q_\mu\,F_0^{BN}(q^2)\right]\,, \non \\
 \la  K(p)| \bar b \ga_{\mu} q|B(p_{B})\ra &=&
 \left[\left(P_\mu- \frac{m^{2}_{B}-m^{2}_{K}}{q^2}\,q_
 \mu\right) F_1^{BK}(q^2) \right.
 \left.+ \frac{m^{2}_{B}-m^{2}_{K}}{q^2}
 \,q_\mu\,F_0^{BK}(q^2)\right]\,, \non \\
 \la K^*(p,\ve_{K^*} )| \bar b \ga_\mu \ga_5 s | B(p_{B})\rangle
&=&i\left\{2m_{K^*}A^{BK^*}_{0}(q^{2})\frac{\ve_{K^*} ^{*}\cdot q}{
q^{2}}q_{\mu
} \right.\non \\
&&+(m_{B}+m_{K^*}) A^{BK^*}_{1}(q^{2})\Big(\ve_{K^{*}\mu }^{*}
-\frac{\ve_{K^*}^{*}\cdot q}{q^{2}}q_{\mu }\Big)\nonumber\\
&&\left.-A^{BK^*}_{2}(q^{2})\frac{\ve_{K^*} ^{*}\cdot q}{m_{B}+m_{K^*}}\Big( P_{\mu }-%
\frac{P\cdot q}{q^{2}}q_{\mu }\Big)\right\}\, .
 \ed
 $A^{BK^*}_{1}(q^{2})$ and $A^{BK^*}_{2}(q^{2})$ are two other form factors that are not relevant in our analysis.
With $\mu=2.0$ GeV, $V_{us}=0.22$, $V_{ub}=3.6\times 10^{-3}
e^{-i\phi_3}$, $\phi_3=72^{\circ}$, $m_N=1.47$ GeV, $m_S=1.50$ GeV
\cite{cheng-chua-liu}, $F^{BN}(m^2_K)=0.26$,
$F^{BN}(m^2_{K^*})=0.28$, $F^{BK}(m^2_S)=0.38$,
$A^{BK^*}_0(m^2_{S})=0.42$ \cite{LF1}, and $f_S=-280$ MeV
\cite{CCY}, one has the following estimation
 \be
A_{KN}+A_{KS}&\approx &\frac{G_F m^3_{B}}{\sqrt{2}} V_{ts} V^*_{tb} \,
\left(-8.50 + 1.37\,i\right)\times 10^{-5}\,,\\
A_{K^*N}+A_{K^*S}&\approx &\frac{G_F m^3_{B}}{\sqrt{2}} V_{ts}
V^*_{tb} \, \left(  1.17 + 0.19\,i\right)\times
 10^{-4}\,.
 \ed
Comparing these values to those coming from the contribution of purely gluonic state
given in Table~\ref{tab:ampm},
one can conclude that the  $q\bar q$ quarkonia contributions
can be safely ignored.

In summary, we have studied the scalar glueball production in
exclusive $B$ decays by using PQCD factorization approach. The
typical momenta carried by the exchanged gluons in the process is
about half of the $B$ meson mass. One thus expects our perturbative
results are trustworthy.  According to our analysis, we find that
the branching fraction for $B^{+}\to K^+ G_{0}$ is a few  $\times 10^{-6}$;
however, for $B^{+}\to K^{*+} G_{0}$ it can be as large as $3 - 4
\times 10^{-5}$. As a result, the branching fraction for the
decaying chain $B^{+}\to K^{(*)+} G_0\to K^{(*)+} (K \bar K)_{G_0}$
is $\sim 0.66( 7.79) \times 10^{-6}$.
With the experimental inputs of  Eqs.(\ref{etaetatokk}) and (\ref{pipitokk}),
we also expect the branching ratios for $B^+ \to K^{(*)+}(\eta\eta)_{G_0}$ and
$B^+ \to K^{(*)+}(\pi\pi)_{G_0}$ are about 50\% and less than 10\% of
$B^+ \to K^{(*)+}(K\bar K)_{G_0}$ respectively.
In this work, we have focused
on the charged $B$ mesons. Similar conclusions can be drawn for the
neutral $B$ mesons where the only difference is their lifetimes.
Their mass difference $(m_{B^0} - m_{B^+})$ is merely $0.33 \pm
0.28$ MeV \cite{pdg} and will not affect our numerical results
significantly. Thus dividing the branching fractions given in Table
\ref{tab:Br} and Table \ref{tab:Br_f} by the ratio
$\tau_{B^+}/\tau_{B^0} = 1.071 \pm 0.009$  \cite{pdg} from direct
measurements, one would obtain the corresponding branching fractions
for the neutral $B$ meson modes.
Experimentally, 
the mode $B^\pm \to (K \bar K)_{f_0(1710)} K^\pm$ has been detected at B{\tiny A}B{\tiny AR} with a branching ratio 
consistent with our PQCD prediction.
This work suggests that  detection of the resonant
three-body mode $B\to (K \bar K)_{f_0(1710)} K^*$ with a predicted larger branching ratio 
can give further support of $f_0(1710)$ is a pure scalar glueball.




\appendix

\section{ Decay amplitude function ${\cal M}_M$ with $k_{\perp} = 0$ \label{ap2}}

Since the mass of glueball is much larger than those of ordinary
pseudoscalars, we find that the influence of transverse momentum on the two-body
decay $B\to K^{(*)} G_0$
is not as large as in the case of $B$ decays into two lighter mesons.
Setting ${\vec k}_{1 \perp}$ and ${\vec k}_{2 \perp}$
in the momenta of the spectator quarks in Eq.(\ref{scalar-meson-momenta})
 %
 to be zero,
 the decay amplitude function ${\cal M}_M$ given in Eq.(\ref{eq:ampm})
 can be simplified to be
\be
  {\cal M}_{M}&=&\frac{f_0 C_{F}}{\pi m_{B}}  \int^{1}_{0}  dx_1 \int^1_0 dx_2 \; \phi_{B}(x_1) \non\\
  && \times  \left\{
      e^{(1)}_{M} \phi_{M}(x_2) + e^{(2)}_{M} \phi^{p}_{M}(x_2)
      +e^{(3)}_{M}
      \phi^{\sigma}_{M}(x_2)
  \right\}\alpha_{s}(t)h(x_1,x_2)\,, \label{eq:ampmnokt}
 \ed
with the hard function $h(x_1,x_2)$ given by
  \be
   h(x_1,x_2)=\frac{1}{x_{1}
   (1-x_1)(r^2_{G_0}+(1-r^2_{G_0})x_2)}\,.
  \ed


\section{Distribution Amplitudes for $K^{(*)}$ \label{DA}}

In this appendix, we compile the light-cone distribution amplitudes that
entered in our calculations.
The distribution amplitudes for $K$, defined in Eq.~(\ref{eq:DA}),
are expressed as follows \cite{BBL_JHEP}:
\be
 \phi_{K}(x)&=&\frac{f_K}{2\sqrt{2N_c}} 6x(1-x)
 \left[1+a^{K}_{1}C^{3/2}_{1}(\xi) + a^{K}_{2} C^{3/2}_{2}(\xi)\right] \,, \non\\
\phi^p_{K}(x)&=&\frac{f_K}{2\sqrt{2N_c}} \left[ 1+3 \rho^{K}_{+}
(1+6a^{K}_2)-9\rho^{K}_{-} a^{K}_{1} + C^{1/2}_{1}(\xi)
 \left( \frac{27}{2} \rho^{K}_{+} a^{K}_{1}-\rho^{K}_{-}
 \left( \frac{3}{2} +27a^{K}_{2}\right)\right) \right.\non\\
&& + C^{1/2}_{2}(\xi) \left(30\eta_{3K} +15 \rho^{K}_{+} a^{K}_2 -3
\rho^{K}_{-} a^{K}_1 \right)+C^{1/2}_{3}(\xi)\left( 10\eta_{3K}
\lambda_{3K} -\frac{9}{2} \rho^{K}_{-}
a^{K}_2\right)\non\\
&& -3 \eta_{3K} \omega_{3K} C^{1/2}_{4}(\xi) +\frac{3}{2} \left(
\rho^{K}_{+}+ \rho^{K}_{-}\right)\left( 1-3a^{K}_1 +6
a^{K}_2\right)\ln(1-x)\non\\
&& \left.+\frac{3}{2} \left(\rho^{K}_{+} - \rho^{K}_{-}
\right)\left(1+3a^{K}_1 +6a^{K}_{2}\right)\ln x
 \right]\,, \non\\
\phi^{\sigma}_{K}(x)&=& \frac{f_K}{2\sqrt{2N_c}} \left\{\xi\left[
b_1 + b_2 C^{3/2}_{1}(\xi) +b_3 C^{3/2}_2(\xi) + b_4
C^{3/2}_{3}(\xi) \right. \right. \non\\
&& \left.\left. -30b_3 x(1-x)+b_5\ln(1-x)+b_6 \ln x \,\right. \right]  \non \\
&& \left. +x(1-x)\left[ -6b_2 +5 b_4 \left(
-21(1-2x)^2+3\right)\right] +\frac{1}{6} \left(-x b_5 +(1-x)b_6
\right) \right\}\,,
 \ed
where $\xi=1-2x$ and the Gegenbauer Polynomials $C^{\nu}_{n}$ are
given by,
 \be
C^{1/2}_{1}(t)&=&t\,, \ \ \ C^{1/2}_{2}(t)=\frac{1}{2}(3t^2-1)\,, \ \ \
C^{1/2}_{3}(t)=\frac{3}{2}\left( \frac{5}{3} t^3-t\right)\,,\non\\
C^{1/2}_{4}(t)&=&\frac{1}{8}\left(3-30t^2+35t^4\right)\,,\non\\
C^{3/2}_{1}(t)&=&3t\,,\ \ \   C^{3/2}_{2}(t)=\frac{3}{2}(5t^2-1)\,,\
\ \ C^{3/2}_{3}(t)=\frac{5}{2}\left(7t^3-3t \right)\,.
 \ed

\noindent
The coefficients $\{b_i\}$ are defined as
 \be
 b_1&=&1+\frac{3}{2} \rho^{K}_{+} + 15 \rho^{K}_{+} a^{K}_2
 -\frac{15}{2}\rho^{K}_{-} a^{K}_{1}\,,\ \ \  b_2= 3\rho^{K}_{+}
 a^{K}_{1} -\frac{15}{2} \rho^{K}_{-} a^{K}_{2}\,,\non\\
 b_3&=& 5\eta_{3K}-\frac{1}{2} \eta_{3K} \omega_{3K} +\frac{3}{2}
 \rho^{K}_{+} a^{K}_{2}\,, \ \ \ b_4=\eta_{3K} \lambda_{3K}\,,
 \non\\
 b_{5(6)} &=& 9 \left(\rho^{K}_{+} \pm \rho^{K}_{-} \right)\left(
1\mp 3a^{K}_1 + 6a^{K}_{2} \right)\,, \ \ \
\rho^{K}_{+}=\frac{(m_s + m_q)^2}{m^2_K} \,, \ \ \
\rho^{K}_{-}=\frac{m_s^2 - m_q^2}{m^2_K}
 \ed
with $m_q$ being the mass of $m_u$ or $m_d$ since $m_u\approx m_d$ is assumed.
Since $m_q\ll m_s$, in our numerical estimations, we take
$\rho^{K}_{+}=\rho^{K}_{-}=\rho^{K}$.
We display the values of decay constant, mass of strange quark, and
relevant coefficients of distribution amplitudes for $K$ meson at $\mu=1$ GeV
in Table~\ref{tab:para_K}.
\begin{table}[hptb]
\caption{ The decay constant, mass of strange quark (in units of
MeV) and coefficients of distribution amplitudes for $K$ meson at $\mu=1$
GeV. }\label{tab:para_K}
\begin{ruledtabular}
\begin{tabular}{cccccccc}
 $f_{K}$& $m_s$ & $a^{K}_{1}$ & $a^{K}_{2}$& $\rho^{K}$ &
$\eta_{3K}$ & $\omega_{3K}$ & $\lambda_{3K}$
 \\ \hline
  160 & 120 & 0.06 & 0.25 & $0.076$ & 0.016 & $-1.2$ &  $1.6$  \\

\end{tabular}
\end{ruledtabular}
\end{table}

Similarly, the distribution amplitude for $K^*$ can  be expressed as
\cite{BZ_PRD71,BBKT}
 \be
\phi_{K^*}(x)&=& \frac{f_{K^*}}{2\sqrt{2N_c}} 6x(1-x)\left[1+3
a^{\parallel}_1 \xi + 3 a^{\parallel}_2 C^{3/2}_{2}(\xi)\right]\,, \non\\
\phi^{p}_{K^*}(x)&=& \frac{f^{T}_{K^*}}{2\sqrt{2N_c}}\left[
3\xi^2+3a^{\perp}_{1} C^{1/2}_{2}(\xi) + a^{\perp}_{2}
C^{3/2}_{2}(\xi)+ 70 \zeta^{T}_{3} C^{1/2}_{4}(\xi) \right.\non\\
&&
\left.+\frac{3}{2}\delta_{+}\left(1+\xi\ln\left( \frac{x}{1-x} \right) \right)
+\frac{3}{2}\delta_{-}\xi \left(2+\ln\left(1-x \right)+\ln
x \right)  \right] \,, \non\\
\phi^{\sigma}_{K^*}(x)&=&
\frac{f^{T}_{K^*}}{4\sqrt{2N_c}}\left\{6\xi
\left[1+a^{\perp}_{1}\xi+ \left(\frac{1}{4} a^{\perp}_{2}
+\frac{35}{6} \zeta^{T}_{3}\right) \left(-20x(1-x)+ 5\xi^2-1
\right)\right] \right.
\non\\
&&\biggl. -12 a^{\perp}_{1}x \left(1-x \right)+3 \delta_{+} \left( 3\xi-2\ln \left(1-x\right)
-2\right) \biggr\}\,.
 \ed
The values of the decay constants and relevant coefficients of the
distribution amplitudes for the $K^*$ meson are shown in Table~\ref{tab:para_Ks}.
\begin{table}[hptb]
\caption{ The decay constants (in units of MeV) and coefficients of
distribution amplitudes for  $K^*$ meson at $\mu=1$ GeV. }\label{tab:para_Ks}
\begin{ruledtabular}
\begin{tabular}{cccccccc}
 $f_{K^*}$ & $f^{T}_{K^*}$ & $a^{\parallel(\perp)}_{1}$ &
$a^{\parallel}_{2}$ & $ a^{\perp}_{2}$ & $\zeta^{T}_3$ & $\delta_+$
& $\delta_{-}$
 \\ \hline
  210 & 170 & 0.10 & 0.09 & 0.13 & 0.024& 0.24 & $-0.24$\\

\end{tabular}
\end{ruledtabular}
\end{table}

{\bf  Acknowledgments}

This work is supported in part by the National Science Council of
R.O.C. under Grant Nos.
NSC-95-2112-M-006-013-MY2 and
NSC-95-2112-M-007-001 and
by the National Center for Theoretical Sciences.


\begin{thebibliography}{99}

\bibitem{scalar-mesons}
N. Mathur, A. Alexandru, Y. Chen, S. J. Dong, T. Draper, I. Horvath, F. X. Lee, K. F. Liu, S. Tamhankar, and
J. B. Zhang, arXiv:hep-ph/0607110.

\bibitem{cheng-chua-liu}
H. Y. Cheng, C. K. Chua, and K. F. Liu,
Phys. Rev. D 74 (2006) 094005 [arXiv:hep-ph/0607206].

\bibitem{he-li-liu-zeng}
X. G. He, X. Q. Li, X. Liu, and X. Q. Zeng, Phys. Rev. D 73 (2006) 051502,
{\it ibid.} D 73 (2006) 114026.

\bibitem{close-zhao}
F. E. Close and Q. Zhao, Phys. Rev. D 71 (2005) 094022.

\bibitem{giacosa-etal}
F. Giacosa, Th. Gutsche, V. E. Lyubovitskij, and A. Faessler, Phys. Rev. D 72 (2005) 094006.

\bibitem{burakovsky-page}
L. Burakovsky and P. R. Page, Phys. Rev. D  59 (1998) 014022.

\bibitem{fariborz}
A.~H.~Fariborz, Phys. Rev. D 74 (2006) 054030
[arXiv:hep-ph/0607105].

\bibitem{chanowitz-talk}
M. Chanowitz, Int. J. Mod. Phys. A21 (2006) 5535 [arXiv:hep-ph/0609217].

\bibitem{lattice}
Y. Chen, A. Alexandru, S. J. Dong, T. Draper, Horv‡th, F. X. Lee, K. F. Liu, N. Mathur, C. Morningstar,
M. Peardon, S. Tamhankar, B. L. Yang, and J. B. Zhang,
Phys. Rev. D 73  (2006) 014516 [arXiv:hep-lat/0510074].

\bibitem{chanowitz}
M.~S.~Chanowitz,
Phys. Rev. Lett.  95 (2005) 172001 [arXiv:hep-ph/0506125].

\bibitem{closeetal}
C. Amsler and F. E. Close, Phys. Lett. B 353 (1995) 385 [arXiv:hep-ph/9505219],
Phys. Rev. D 53 (1996) 295 [arXiv:hep-ph/9507326];
F. E. Close, G. R. Farrar, and Z.-p. Li, Phys. Rev. D 55 (1997) 5749 [arXiv:hep-ph/9610280].

\bibitem{he-jin-ma}
X. G. He, H. Y. Jin, and J. P. Ma,
Phys. Rev. D 66 (2002) 074015 [arXiv:hep-ph/0203191].

\bibitem{zhao-close}
Q. Zhao and F. E. Close,
Int. J. Mod. Phys. A 21 (2006) 821 [arXiv:hep-ph/0509305].

\bibitem{brodsky}
S. Brodsky, A. S. Goldhaber, and J. Lee,  Phys. Rev. Lett.  91 (2003) 112001.

\bibitem{he-yuan}X.~G. He and T.~C. Yuan,
arXiv:hep-ph/06121082.

\bibitem{pdg}
Review of Particle Physics, Particle Data Group, J. Phys. G: Nucl. Part. Phys. 33 (2006) 1.


\bibitem{BBL} G. Buchalla, A. J. Buras, and M. E. Lautenbacher, Rev.
Mod. Phys.  68 1125 (1996).


\bibitem{he-lin}
X.~G.~He and G. L.~Lin,
Phys. Lett. B 454 (1999) 123.

\bibitem{GFA} Y.~H. Chen, H.~Y. Cheng, B. Tseng, and K.~C. Yang, Phys. Rev D 60
(1999) 094014.

\bibitem{Bjorken} J. D. Bjorken, Nucl. Phys. Proc. Suppl.  11 (1989) 325.


\bibitem{PQCD1}
G. P. Lepage and S. J. Brodsky, Phys. Lett. B 87 (1979) 359, Phys. Rev. D 22 (1980) 2157;
H. N. Li and G. Sterman, Nucl. Phys. B 381 (1992) 129;
G. Sterman, Phys. Lett. B 179 (1986) 281, Nucl. Phys. B 281 (1987) 310;
S. Catani and L. Trentadue, Nucl. Phys. B 327 (1989) 323, Nucl. Phys. B 353 (1991) 183.

\bibitem{PQCD2}
T. W. Yeh and H. N. Li, Phys. Rev. D 56 (1997) 1615;
H. N. Li, Phys. Rev. D 64 (2001) 014019;
H. N. Li, Phys. Rev. D 66 (2002) 094010.

\bibitem{GN_PRD55}
A. G. Grozin and M. Neubert, Phys. Rev. D 55 (1997) 272.

\bibitem{K-meson}
V. M. Braun and I. E. Filyanov, Z Phys. C 48 (1990) 239;
P. Ball, JHEP 01 (1999) 010.

\bibitem{KLS_PRD65}
T. Kurimoto, H. N. Li, and A. I. Sanda, Phys. Rev. D 65 (2002) 014007.

\bibitem{BN_NPB651} M. Beneke and M. Neubert, Nucl. Phys. B
651(2003)  225.

\bibitem{Li}
 Y.~Y.~Charng, T.~Kurimoto, and H.~N.~Li,
  Phys. Rev. D 74  (2006) 074024.

\bibitem{CL_PRD63}
C. H. Chen and H. N. Li, Phys. Rev. D 63 (2000) 014003.

\bibitem{LL_PRD70} H.~N. Li and H.~S. Liao, Phys. Rev. D 70 
(2004) 074030.

\bibitem{CS} J.~C. Collins and D.~E. Soper, Nucl. Phys. B 193 (1981) 381.

\bibitem{BS} J. Botts and G. Sterman, Nucl. Phys. B  325 (1989) 62.

\bibitem{BBL_JHEP}
P. Ball, V. M. Braun, and A. Lenz, JHEP 05 (2006) 004.

\bibitem{BZ_PRD71}
P. Ball and R. Zwicky, Phys. Rev. D 71 (2005) 014029.

\bibitem{BBKT}
P. Ball, V. M. Braun, Y. Koike, and K. Tanaka, Nucl. Phys. B 529 (1998) 323.


\bibitem{chao-he-ma}
K. T. Chao, X. G. He, and J. P. Ma, hep-ph/0512327.

\bibitem{flavor0}
 J.~Schechter, A.~Subbaraman, and H.~Weigel,
  Phys. Rev. D 48 (1993) 339.

\bibitem{flavor}
T. Feldmann, P. Kroll, and B. Stech, Phys. Rev. D 58, (1998) 114006;
Phys. Lett. B 645 (2007) 197;
A. G. Akeroyd, C. H. Chen, and C. Q. Geng, Phys. Rev. D 75 (2007) 054003.

\bibitem{KLOE} F. Ambrosino {\it et al.} (KLOE Collaboration),
arXiv:hep-ex/0612029.

\bibitem{CCS} H. Y. Cheng, C. K. Chua, and A. Soni, arXiv:0704.1049; 
Phys. Rev. D 72 (2005) 094003.

\bibitem{chao-he-ma-comment}
K. T. Chao, X. G. He, and J. P. Ma, 
Phys. Rev. Lett. 98 (2007) 149103 [arXiv:hep-ph/0704.1061].

\bibitem{chanowitz-reply}
M. Chanowitz, Phys. Rev. Lett. 98 (2007) 149104 [arXiv:hep-ph/0704.1616].


\bibitem{babar-br}
B. Aubert {\it et al.} (B{\tiny A}B{\tiny AR} Collaboration), Phys. Rev. D 74 (2006) 032003.


\bibitem{LW} W. Lee and D. Weingarten, Phys. Rev. D 61 (1999)
014015.

\bibitem{CG_PRD75} C. H. Chen and C. Q. Geng, Phys. Rev. D 75 (2007) 054010.

\bibitem{LF1} H.~Y. Cheng, C.~K. Chua, and C.~W. Hwang, Phys. Rev. D 69 (2004) 074025.

\bibitem{CCY} H. Y. Cheng, C. K. Chua, and K. C. Yang, Phys. Rev. D 73 (2006) 014017.

\end{thebibliography}
\end{document}